\documentclass[%
 reprint,
 amsmath,amssymb,
 aps,
prb,
]{revtex4-1}

\usepackage{graphicx}
\usepackage{dcolumn}
\usepackage{bm}

\begin{document}

\preprint{APS/123-QED}

\title{Gyrotropic skyrmion modes in ultrathin magnetic circular dots}
\author{Konstantin Y. Guslienko,$^{1,2}$ Zukhra V. Gareeva,$^{3}$}%
\affiliation{$^1$Depto. Fisica de Materiales, Facultad de Quimica, Universidad del Pais Vasco, UPV/EHU, 20018 San Sebastian, Spain}
\affiliation{$^2$IKERBASQUE, the Basque Foundation for Science, 48013 Bilbao, Spain}
\affiliation{$^3$Institute of Molecule and Crystal Physics, Russian Academy of Sciences, 450075 Ufa, Russia}
\date{\today}

\begin{abstract}
We calculate low-frequency gyrotropic spin excitation modes of the skyrmion ground state ultrathin cylindrical magnetic dots. The skyrmion is assumed to be stabilized at room temperature and zero external magnetic field due to an interplay of the isotropic and Dzyaloshinskii-Moriya exchange interactions, perpendicular magnetic anisotropy and magnetostatic interaction. We consider Bloch- and Neel-type magnetic skyrmions and assume that the dot magnetization does not depend on the thickness coordinate. The skyrmion gyrotropic frequencies are calculated in GHz range as a function of the skyrmion equilibrium radius, dot radius and the dot magnetic parameters. Recent experiments on magnetic skyrmion gyrotropic dynamics in nanodots are discussed.

\begin{description}

\item[PACS numbers]
75.75.+a,75.60.Jk,75.30.Gw
\end{description}
\end{abstract}
\maketitle

\section{Introduction}

Magnetic skyrmions received recently considerable attention due to their unusual physical properties and promising applications in nanoelectronics, information data storage, spintronics etc. \cite{fert2013skyrmions,sampaio2013nucleation}. Nanoscale dimensions, considerable stability, high mobility, several controllable parameters (polarity, chirality, topological charge) make them attractive for the implementations in information storage and processing devices on nanoscale.

Being a kind of magnetic topological solitons \cite{kosevich1990magnetic} in 2D spin systems, skyrmions exhibit a wide variety of unusual properties that are related to their non-trivial topology. The first theoretical estimation of the stability of localized topological solitons in infinite ferromagnets has been presented by Dzyaloshinskii et al. \cite{dzyloshinskii1979localized} Then, it was noticed \cite{bogdanov1989thermodynamically} that the Dzyaloshinskii -– Moriya exchange interaction (DMI) stabilizes 2D magnetic vortices (skyrmions in modern terminology) in magnetic systems whose symmetry group lacks of the space inversion symmetry operation. Twenty years later 2D hexagonal skyrmion lattices were experimentally detected in magnetic films with bulk (cubic B20 compounds, like MnSi, FeGe) \cite{muhlbauer2009skyrmion,yu2010real,huang2012extended}, and interfacial types \cite{sonntag2014thermal,heinze2011spontaneous} of DMI. It is accepted that the bulk DMI stabilizes Bloch skyrmions in B20 magnets \cite{muhlbauer2009skyrmion,yu2010real,huang2012extended}, whereas the interfacial DMI leads to formation of Neel skyrmions \cite{sonntag2014thermal} in ultrathin multilayer films. The B20 skyrmions are stable predominantly at low temperatures and finite magnetic fields \cite{muhlbauer2009skyrmion,yu2010real,huang2012extended,sonntag2014thermal}. However, metastable skyrmion lattices can also exist at zero magnetic fields in thin epitaxial FeGe films\cite{huang2012extended} and monolayer Fe films on Ir(111) surface \cite{heinze2011spontaneous}.

Just few years ago simulations \cite{sampaio2013nucleation} and experiments \cite{woo2015observation,moreau2015skyrmions,moreau2016additive,boulle2016room}showed that skyrmions can be stabilized at room temperatures (RT) in ultrathin multilayer structures deposited by sputtering (Co/Pt \cite{woo2015observation,moreau2015skyrmions,moreau2016additive,boulle2016room}, and Ir/Co/Pt \cite{moreau2015skyrmions,moreau2016additive}), including magnetic dots. Moreau-Luchaire et al. \cite{moreau2015skyrmions,moreau2016additive} reported the observation of RT skyrmions in multilayer films Pt/Co/Ir composed of heavy metal and ferromagnetic layers where the single skyrmions are stabilized by chiral DMI. Very recently, RT skyrmions were observed in bulk CoZnMn alloys \cite{tokunaga2015new}, in Fe/Ni/Cu/Ni/Cu multilayer films \cite{chen2015room}, and in tri-layer stripes Ta/CoFeB/TaO \cite{jiang2015blowing}. In the most cases, the existence of magnetic skyrmions is related to DMI. However, the alternative approaches referring to artificial RT skyrmion crystals \cite{sun2013creating,miao2014experimental,li2014tailoring,gilbert2015realization,sapozhnikov2015two}, skyrmions stabilized by perpendicular magnetic anisotropy \cite{guslienko2015skyrmion} without DMI, dynamically stabilized skyrmions \cite{zhou2015dynamically}, etc., have been also developed.

The complex structure of magnetic skyrmions representing particle-like nanosize objects containing thousands of spins results in their reach dynamics. A number of associated extraordinary findings including emergent electromagnetic fields, topological Hall effect, ultralow densities of spin currents driving skyrmion motion, skyrmion breathing and rotation excitation modes etc. were reported in recent years \cite{yu2012skyrmion,mochizuki2012spin,onose2012observation,okamura2013microwave,lin2014internal,schwarze2015universal,ogawa2015ultrafast,mochizuki2015dynamical}.

Low and high frequency spin dynamics over a skyrmion background being of particular interest has been actively explored in 2D skyrmion lattices \cite{mochizuki2012spin,onose2012observation,okamura2013microwave,lin2014internal,schwarze2015universal,ogawa2015ultrafast}. Mochizuki \cite{mochizuki2012spin}showed the existence of two in-plane rotational eigenmodes (clockwise and anticlockwise) of skyrmion spin texture unit cell and one breathing mode (oscillations of the skyrmion radius) in the GHz frequency range. Later on, the experiments detecting internal skyrmion modes have been carried out \cite{onose2012observation,okamura2013microwave,lin2014internal,schwarze2015universal,ogawa2015ultrafast}. However, the microwave measurements \cite{onose2012observation,okamura2013microwave} confirmed the existence of only one of the skyrmion rotating modes and a breathing mode. Contrary, two rotating spin eigenmodes and one breathing mode were detected by time-resolved magneto-optics in Cu$_2$OSeO$_3$ \cite{ogawa2015ultrafast}. Very recent broadband ferromagnetic resonance measurements of MnSi, FeCoSi and Cu$_2$OSeO$_3$ \cite{schwarze2015universal} showed existence of only two excitation modes in the spectra of these skyrmion crystals. Therefore, there is no clear understanding of the B20 skyrmion spin excitation modes to the moment.

To understand the complicated skyrmion spin excitation spectra researches appeal to simplified systems: isolated skyrmions and magnetic bubbles in ferromagnetic films and dots. Additional reason for that are perspective applications of the RT isolated skyrmions in nanoscale devices \cite{zhang2015current}. Isolated skyrmions, conditions required for their existence in patterned films \cite{woo2015observation,moreau2015skyrmions,moreau2016additive,boulle2016room,guslienko2015skyrmion,rohart2013skyrmion,beg2015ground}, and the dynamics of isolated skyrmions in magnetic films and nanodots have been simulated \cite{lin2014internal,zhang2015current,kim2014breathing,gareeva2016magnetic} and experimentally explored \cite{buttner2015dynamics}. Lin et al. \cite{lin2014internal} simulated dynamics of isolated skyrmion in an infinite film and found one gyrotropic mode, several breathing and non-symmetric skyrmion shape modes in the area of the skyrmion stability. The skyrmion gyrotropic mode excited by the spin polarized current in a nanopillar was simulated in Ref. \cite{zhang2015current}. Simulations of the skyrmion breathing modes in cylindrical dots were presented in Ref. \cite{kim2014breathing}.
\linebreak
\indent
Spin excitation spectrum in presence of the magnetic vortex (half-skyrmion) ground state was investigated in detail in Ref. \cite{guslienko2008magnetic}. It includes a low-frequency gyrotropic mode (sub-GHz range) and high-frequency radially/azimuthally symmetric spin waves classified by integer indices - number of nodes of the dynamical magnetization in the radial/azimuthal directions \cite{guslienko2008magnetic}. Similar classification of the spin eigenmodes can be applied to isolated skyrmions in magnetic dots and infinite films. The spin waves on the Bloch skyrmion background were recently calculated in Ref. \cite{gareeva2016magnetic} within a simple model. In general, the skyrmion magnetization oscillations can be represented as a superposition of spin eigenmodes –- spin waves and gyrotropic modes. Gyrotropic modes correspond to translation motion of the skyrmion center around its equilibrium position. There are also standing and travelling spin waves with quantized eigenfrequencies in a confined system. The frequencies of excited spin eigenmodes depend on a variety of factors (system geometry (e.g., nanodot diameter), magnetic anisotropy, exchange etc.).\linebreak
\indent The experimental investigations of spin dynamics of the RT isolated skyrmions are in a beginning stage. Sometimes it appears to be difficult to recognize experimentally the type and origin of a particular skyrmion excitation mode and a theoretical input is crucial. Two spin excitation modes, presumably gyrotropic ones, of magnetic bubble skyrmions in CoB/Pt multilayer dots with CoB layer thickness of $0.4$ nm were recently measured by Buettner et al. \cite{buttner2015dynamics}. They detected clockwise (CW) and counterclockwise (CCW) rotating spin modes, and an unusual pentagon like trajectory of the bubble skyrmion core.\\
\indent In this article we focus on skyrmion gyrotropic excitation modes considering the skyrmion as a ground state of thin cylindrical magnetic dot. We calculate the mode eigenfrequencies within rigid skyrmion model accounting the exchange, DMI, magnetostatic interaction and uniaxial magnetic anisotropy. We show that for the given dot parameters there exists just one gyrotropic mode, which can be CCW or CW depending on the skyrmion core magnetization direction.\\
\section{Theory}
For consideration of magnetic skyrmion dynamics in the dots we use theoretical approach based on the Landau-Lifshitz (LL) equation of magnetization $(\bm M )$ motion, $\dot{\bm M}=-\gamma \bm M \times \bm H_{eff}$, where $\bm H_{eff}=-\delta w/ \delta \bm M$, $\gamma$ is the gyromagnetic ratio and $w$ is the magnetic energy density
\begin{eqnarray}
 \label{eq:A1}
&&w=A(\partial_\mu m_\alpha)^2+w_D-\kappa m_z^2-\frac{1}{2}M_s\bm m\cdot\bm H_m
\end{eqnarray}			
where $A$ is the isotropic exchange stiffness constant proportional to the Heisenberg exchange integral, $K>0$ is the constant of uniaxial magnetic anisotropy, $\alpha, \mu =x, y, z$,  $\bm m=\bm M/M_s$ is the unit magnetization vector, $M_s$ is the saturation magnetization, and $\bm H_m$ is the magnetostatic field.
The DMI term in Eq. (\ref{eq:A1}) can be represented in two forms: 1. $w_D=D(\bm m\cdot rot\bm m)$  \cite{fert2013skyrmions,sonntag2014thermal,yu2012skyrmion,mochizuki2012spin,onose2012observation,okamura2013microwave,lin2014internal,schwarze2015universal,ogawa2015ultrafast} and  2. $w_D^*=D[m_z(\nabla \cdot \bm m)-(\bm m\cdot \nabla)m_z]$, see Refs. \cite{sampaio2013nucleation,woo2015observation,moreau2015skyrmions,moreau2016additive,boulle2016room,chen2015room,buttner2015dynamics} and references therein. Here $D$ is the constant of the Dzyaloshinskii – Moriya exchange interaction. The first form is used for bulk DMI in the B20 cubic compounds (MnSi, FeGe, etc.), and the second is used for interfacial induced DMI in the case of an interface of ultrathin metallic ferromagnet with a nonmagnetic metal having a strong spin orbit coupling \cite{sampaio2013nucleation}, i.e., Co/Pt, Co/Pd. The DMI term  $w_D^*$ favors to stabilization of so-called $\phi-$skyrmions, where the skyrmion core and peripheries are separated by a Bloch domain wall (Fig. 1a). The interface DMI term   leads to stabilization of $\rho-$skyrmions, where the skyrmion core and peripheries are separated by a Neel domain wall (Fig. 1b). The DMI is not of principal importance for $\phi-$skyrmions in dots, which can be stable even at $D=0$ \cite{guslienko2015skyrmion,buttner2015dynamics}. However, the interface DMI essentially contributes to the $\rho-$skyrmion stabilization and dynamics, the skyrmions are not stable if the DMI strength is lower than some critical value \cite{sampaio2013nucleation,rohart2013skyrmion}.
\begin{figure}
\includegraphics[width=86mm]{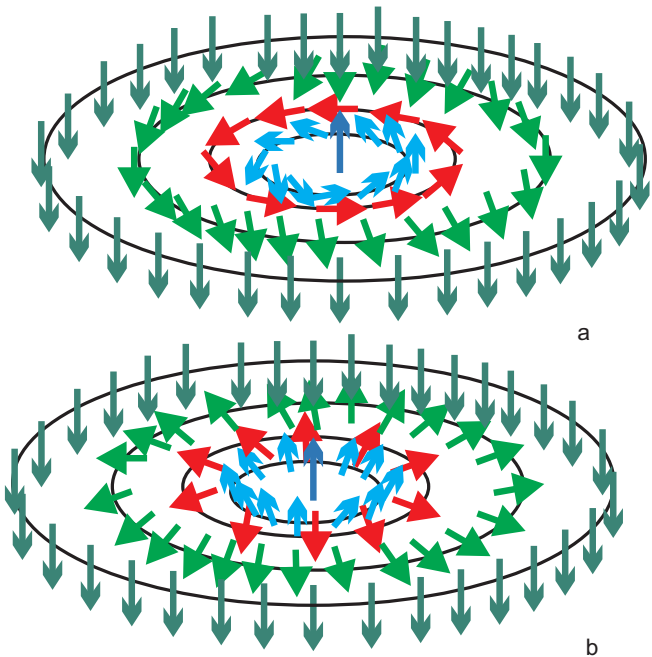}
\caption{\label{fig:Fig1} Magnetic skyrmion textures: a) Bloch ($\phi$) skyrmion (magnetization rotates in the plane perpendicular to the radial direction), b) Neel $(\rho)$ skyrmion (magnetization rotates along the in-plane radial direction).}
\end{figure}

It is convenient to rewrite the LL equation of motion of the dot magnetization $\bm M(\bm \rho,t)=\bm M(\bm \rho,\bm X(t))$ in the form of Thiele equation of motion for the skyrmion core position  $\bm X=(X,Y)$ in the dot via complex variables $s=s_x+is_y$, $\bm s=\bm X/R$:
\begin{eqnarray}
 \label{eq:M1}
&&iG_z\dot{s}=\frac{2}{R^2}\frac{\partial W}{\partial \bar{s}},
\end{eqnarray}
 where $\partial/\partial \bar{s}=(\partial/\partial s_x+i\partial/\partial s_y)/2$, $G_z=-p|G|$ are the $z$-projection and absolute value of the gyrovector, $L$ is the dot thickness, $R$ is the dot radius, $W=L\int{d^2\bm \rho w}$  is the total dot magnetic energy,  $p=\pm 1$ is the skyrmion core polarization. The skyrmion magnetic energy can be decomposed in series on small parameter  $|s|<<1$ as $W(s)=W(0)+\kappa |s|^2/2$, where $\kappa$ is the stiffness coefficient.
Let us use the cylindrical coordinates $(\rho,\phi)$ to describe the in-plane radius vector $\bm \rho(x,y)$. We assume that the dot is thin enough, so there is no dependence of the skyrmion magnetization on the thickness $z$- coordinate and the $\phi-$ and $\rho-$ skyrmion ground states are clearly defined. The problem is reduced to calculation of the skyrmion magnetic energy $W(s)$ as a function of the skyrmion displacement $s$. We use complex variables $z=(x+iy)/R$ and an analytic function $f(z)$ to describe the skyrmion dynamics in 2D ferromagnet. The linearized system of Eqs. (\ref{eq:M1}) has the same form for the $\phi-$skyrmion and $\rho-$skyrmion backgrounds if the magnetostatic energy related to the volume and side surface charges is neglected.
The magnetization components can be expressed as
\begin{eqnarray}
 \label{eq:M2}
&&m_x+im_y=\frac{2f(z)}{1+|f(z)|^2}\nonumber\\
&&m_z=\frac{1-|f(z)|^2}{1+|f(z)|^2}
\end{eqnarray}
 The isotropic exchange and DMI energy densities are written in the form
 \begin{eqnarray}
 \label{eq:M3}
&&w_{ex}=\frac{1}{R^2}\frac{8A}{(1+|f(z)|^2)^2}|\frac{\partial f}{\partial z}|^2\nonumber\\
&&\bar w_{D}=\frac{1}{R}\frac{4D}{(1+|f(z)|^2)^2} \frac{\partial f}{\partial z}
\end{eqnarray}
Then, the interface (bulk) DMI densities are equal to $w_D^*=Re(\bar w_{D})$  or  $w_D=Im(\bar w_{D})$, respectively. The typical DMI parameter $D$ is about of 1 $mJ/m^2$.
Accounting ultrathin dot thickness about 1 $nm$ ($L<<R$) we apply the rigid skyrmion approximation  $f(z)=e^{i\Phi_0}(z-s)/c$, which is asymptotically exact within the limit of infinite film with dominating exchange interaction, $L/R \rightarrow 0$, and corresponds to the skyrmion equilibrium profile  $\cos{\Theta_0}=(R_c^2-\rho^2)/(R_c^2+\rho^2)$. Here, $c=R_c/R$  is the reduced skyrmion radius, and the skyrmion phase is  $\Phi_0=C\pi/2$ ( $C=\pm 1$ is the skyrmion chirality) for $\phi$-skyrmion or $\Phi_0=0, \pi$ for $\rho$-skyrmion. To unify description of $\phi$- and $\rho$-skyrmions we define the generalized skyrmion chirality $C$ as follows. $C=sign(m_{\phi})=\sin(\Phi_0)$ for $\phi$-skyrmions and $C=sign(m_{\rho})=\cos(\Phi_0)$ for $\rho$-skyrmions. The static DMI energy density is proportional to the product $DC$, and is minimal at $DC=-|D|<0$ defining the chiral skyrmion ground state. Nonzero DMI strength lifts the energy degeneracy with respect to the skyrmion chirality $C$.
It is evident that in the exchange approximation (4) the DMI energy just renormalizes the isotropic exchange energy,  $A\rightarrow A'=A+DCR_c/2$.

The exchange contribution to the stiffness coefficient is
\begin{eqnarray}
 \label{eq:M4}
&&\kappa_{ex}(c)=-32\pi A L\frac{c^2}{(1+c^2)^3}
\end{eqnarray}
The anisotropy energy $w_a=-Km_z^2$  in Eq. (\ref{eq:A1}) contribution to the stiffness is calculated to be
 \begin{eqnarray}
 \label{eq:M5}
&&\kappa_{a}(c)=-8\pi K R^2 L\frac{c^2(1-c^2)}{(1+c^2)^3}
\end{eqnarray}
To calculate the skyrmion magnetostatic energy $w_m$  we distinguish the energy of the bulk (non--zero for $\rho-$skyrmions), side surface and face dot surface magnetic charges. The energy cannot be simply expressed via the analytical function $f(z)$, therefore, we used a direct calculation of $w_m$  via the magnetization bulk $div \bm m$ and surface divergence ($\bm m\cdot\bm n$), where the vector $\bm n$ is normal to the dot surface ( $\bm n=\bm \hat{z}, \bm \hat{\rho}$ for the face and side surface charges, correspondingly).
The magnetostatic energy of the face and side surface charges of the displaced skyrmion can be calculated by the equation
 \begin{eqnarray}
 \label{eq:M6}
&&W_m(\bm s)=\frac{1}{2}M_s^2\int{dS\int{dS'\frac{m_n(\bm r,\bm s)m_n(\bm r',\bm s)}{|\bm r - \bm r'|}}}
\end{eqnarray}
where $m_n=(\bm m\cdot\bm n)$ is the surface divergence.
The side surface charges energy is proportional to the function $F(\beta)=\int_0^\infty{dk f(\beta k)J_1^2(k)/k}$, where $J_1(x)$ is the first order Bessel function,  $f(x)=1-(1-exp(-x))/x$, and  $\beta=L/R$. Detailed calculations of the skyrmion magnetostatic energy will be published elsewhere. Here we note that the bulk and side surface charges energies are proportional to the small dot aspect ratio $L/R$ and can be neglected for ultrathin dots. Whereas, the face charges energy main term renormalizes the uniaxial anisotropy constant, $K\rightarrow K-2\pi M_s^2$.
We can write the total stiffness coefficient for ultrathin dot as
\begin{eqnarray}
 \label{eq:M7}
&&\kappa(c)=16\pi M_s^2 R^2 L\frac{c^2}{(1+c^2)^3}[\pi(1-Q)(1-c^2)-\nonumber\\
&&(\frac{1}{r})^2(1+\frac{1}{2} rdc)]
\end{eqnarray}
where  $Q=K/2\pi M_s^2$ is the dot magnetic material quality factor,  $L_e=\sqrt{2}A/M_s$ is the exchange length,$r=R/L_e$ is the reduced dot radius and dimensionless DMI parameter $d=DCL_e/A$ for $\phi$- and $\rho$-skyrmions.
The gyrovector $G_z=-p|G|$ of the centered skyrmion ($s=0$) is proportional to the skyrmion topological charge \cite{guslienko2015skyrmion} and can be calculated using the definition
 \begin{eqnarray}
 \label{eq:M8}
&&G_z=\frac{M_s L}{\gamma}\int{d^2\bm \rho \bm m [\partial_x \bm m \times \partial_y \bm m ]},
\end{eqnarray}
or via the complex variables as
 \begin{eqnarray}
 \label{eq:M9}
&&|G|=\frac{M_s L}{\gamma}\int{d^2\bm \rho \frac{4}{(1+|f|^2)^2}|\frac{\partial f}{\partial z}|^2}\nonumber\\
&&|G(c)|=\frac{M_s L}{\gamma}\frac{4 \pi}{1+c^2}.
\end{eqnarray}
	The skyrmion gyrotropic frequency calculated from the Thiele equation of motion (\ref{eq:M1}) using eqs. (\ref{eq:M8})-(\ref{eq:M9}) is  $\omega(c)=\kappa(c)/|G(c)|R^2$, or explicitly is given by

\begin{eqnarray}
 \label{eq:A2}
&&\omega_G=\omega_M\frac{c^2}{1+c^2}[(1-Q)(1-c^2)-\frac{1}{\pi}\frac{1}{r^2}(1+\frac{1}{2}rdc)],
\end{eqnarray}
where  $\omega_M=4\gamma \pi M_s$, and the reduced equilibrium skyrmion radius $c=R_c/R$  is a function of the dot sizes and magnetic parameters $A, D, K, M_s$.

\section{Results and Discussion}

We calculated a low-frequency skyrmion dynamics of $\rho-$ and $\phi-$skyrmions differing by the type of a domain wall (Bloch and Neel) in the skyrmion spin configuration. An ultrathin circular ferromagnetic dot with radius $R$ about of 100 $nm$ and thickness $L$ about of 1 nanometer was considered. For such small dot thickness the magnetostatic energy is reduced to the magnetic energy of the face charges, which can be accounted in the simplified form of an effective easy-plane anisotropy.

In the main approximation the skyrmion eigenmodes can be conventionally divided into internal (low frequency) and external (high frequency) modes. The internal modes related to weak skyrmion deformations are localized close to the skyrmion center and include translation (gyrotropic) and breathing modes. The high–frequency spin wave modes are delocalized and occupy the whole dot volume. We calculate the skyrmion spin excitation modes that are closely related to the skyrmion topological charge, i.e., translation or gyrotropic modes. These collective spin modes correspond to the skyrmion rotation around an equilibrium position corresponding to minimum of the total magnetic energy. Based on the Thiele collective coordinate approach we describe the motion of skyrmion center position in a magnetic dot within rigid skyrmion model accounting the exchange, DMI, magnetostatic interactions and uniaxial magnetic anisotropy, calculate the frequency of circular skyrmion rotation (gyrotropic freq
 uency) and compare results with recent experimental data and micromagnetic simulations.

The skyrmion gyrotropic frequency of ultrathin cylindrical dots given by Eq. (11) is represented via the skyrmion radius $c$. The equilibrium value of the skyrmion radius $c$ can be found using minimization of the total magnetic energy given in Ref. \cite{guslienko2015skyrmion} within the ultrathin dot limit $L/R\rightarrow 0$  adding the DMI term by the substitution of the exchange stiffness $A\rightarrow A'=A+DCR_c/2$. The gyrotropic frequency (\ref{eq:A2}) is proportional to the inverse in-plane magnetic susceptibility similarly to the vortex gyrotropic frequency \cite{guslienko2008magnetic} and is positive in the stable skyrmion state. Otherwise, the skyrmion will escape from the dot lowering its energy. The line $\omega_G(c)=0$  marks the border of the skyrmion stability within the model. Accounting the reduced skyrmion radius $c<1$ we note that if the DMI parameter $d$ of any sign is small ($|d|<d_c=2L_e/R_c$) the skyrmion in an ultrathin dot is stabilized by the face magnetostatic interaction at
  moderate perpendicular anisotropy $Q < 1$. This is typical for $\phi-$skyrmions \cite{guslienko2015skyrmion,buttner2015dynamics}, but is valid also for $\rho-$skyrmions in ultrathin dots. If the parameter $d$ is negative ($d=-|d|<0$ corresponds to the skyrmion ground state) and its magnitude is large (above the critical value $|d|>d_c$), then the $\rho-$skyrmions might be stabilized by the DMI in ultrathin dots having a large perpendicular anisotropy $Q > 1$ \cite{sampaio2013nucleation,woo2015observation,moreau2015skyrmions,moreau2016additive,boulle2016room}. The gyrotropic frequency $\omega_G$  for such $\rho-$skyrmions decreases approximately as $1/R$ at the dot radius $R$ increasing. We note that the similar dependence $\omega_G(R)\approx 1/R$ was calculated by Guslienko et al. \cite{guslienko2008magnetic} for the vortex gyrotropic frequency in thin cylindrical dots due to the dominating magnetostatic interaction.
The $\phi-$ and $\rho-$skyrmion states are stable in circular magnetic dots within some range of the parameters according to Refs. \cite{sampaio2013nucleation,schwarze2015universal,mochizuki2015dynamical}.
Estimation of the critical value of the DMI parameter $D_c=2A/R_c$ using typical parameters $A$=10 $pJ/m$ and $R_c$ =10 $nm$ yields $D_c$ = 2 $mJ/m^2$ in good agreement with experiments \cite{moreau2015skyrmions,moreau2016additive,boulle2016room}.  X-ray imaging of magnetic skyrmions in the ultrathin multilayer circular dots Co/Pt \cite{woo2015observation,moreau2015skyrmions,moreau2016additive,boulle2016room}, Ir/Co/Pt \cite{moreau2015skyrmions,moreau2016additive} and CoB/Pt \cite{buttner2015dynamics} showed that the skyrmion radius $R_c$  is several tens of nm, much smaller than the dot radius and the typical ratio $c=R_c/R= 0.1-0.2$ is small. This is in some disagreement with micromagnetic simulations of the $\rho-$skyrmion stability in the ultrathin circular dots, where the larger values of $c = 0.3-0.5$ strongly dependent on the DMI strength, dot radius \cite{sampaio2013nucleation}, and dot thickness \cite{boulle2016room} have been found.

The positive skyrmion core polarization $p=m_z(0)$ corresponds to the counter clock-wise (CCW) skyrmion core gyrotropic rotation and negative  $p=-1$ corresponds to the gyrotropic mode rotating clock-wise (CW). I.e., there is always only one gyrotropic mode for the given skyrmion polarization $p$ and corresponding sign of the gyrovector (see Theory). The second low-frequency mode (CW for $p=+1$) simulated by Mochizuki \cite{mochizuki2012spin,mochizuki2015dynamical} cannot be interpreted as a second gyrotropic eigenmode. In the simulations \cite{mochizuki2012spin,mochizuki2015dynamical} the sign of $p$ was determined by the sign of the bias magnetic field perpendicular to the film plane. It was noticed by Mochizuki et al. \cite{mochizuki2015dynamical} that the CCW low-frequency mode has intensity much larger than intensity of the higher-frequency CW mode. I.e., the CCW gyrotropic mode is a sole resonance mode for $p=+1$, or eigenmode of the system.

More precisely, moving skyrmion cannot be considered as an absolutely rigid object, its dynamical profile is deformed that can be represented as a hybridization with azimuthal spin waves excited over the skyrmion background \cite{gareeva2016magnetic} and resulting in a finite skyrmion inertia term in the Thiele equation of motion. The low frequency gyrotropic eigenmodes are closely related to non-zero skyrmion mass \cite{buttner2015dynamics,guslienko2015giant}. Recently two gyrotropic modes in the CoB/Pt circular dots modes were simulated and measured by X-ray imaging technique \cite{buttner2015dynamics}. According to the interpretation of Ref. \cite{buttner2015dynamics} existence of two gyrotropic modes (rotating in opposite directions) corresponds to a finite skyrmion mass. However, the second higher frequency gyrotropic mode can be interpreted as azimuthal spin wave \cite{gareeva2016magnetic,guslienko2008magnetic,guslienko2015giant}, and there is only one gyrotropic mode in the dot spin excitation spectrum. Nevertheless, the skyrmions can have a considerable mass accounting for their magnetic energy change increasing skyrmion velocity. The skyrmion mass might also depend on mechanism driving skyrmion motions (external field gradient, temperature, current, etc.).

We apply Eq. (\ref{eq:A2}) to estimate the gyrotropic frequency of the bubble $\phi-$skyrmion measured in Ref. \cite{buttner2015dynamics}. The dot radius is $R$=275 $nm$, $M_s$ = 1190 $G$, $Q$=0.866, and $D=0$. Using the reduced value of $A$=15 $pJ/m$ typical for ultrathin Co-films \cite{sampaio2013nucleation,moreau2015skyrmions,moreau2016additive,rohart2013skyrmion} we get with these parameters  $\omega_M/2\pi$=43.5 $GHz$ (assuming typical value of $\gamma/2\pi$= 2.9 $MHz/Oe$) and  $L_e$=14.5 $nm$. The total magnetic material thickness is 12 nm (30 repeats of ultrathin CoB layer), and the magnetostatic energy in Eq. (\ref{eq:A2}) should be accounted more precisely. This energy for displaced $\phi-$skyrmion appears due to side surface magnetic charges and can be accounted by adding the term  $F(\beta)(1+c^2)$ (see definition of the function  $F(\beta)$ in Sec. 2) to the square bracket in Eq. (\ref{eq:A2}). The gyrotropic frequency calculated by corrected Eq. (\ref{eq:A2}) is equal to
  1.11 $GHz$, whereas the value of 1.00 $GHz$ was detected for the low frequency CCW mode in Ref. \cite{guslienko2008magnetic}. The agreement is reasonable good accounting that the relaxation time in Ref. \cite{guslienko2008magnetic} is comparable with the magnetization oscillation period. That did not allow determining the skyrmion eigenfrequencies with good accuracy and therefore, the gyrotropic frequencies and skyrmion mass obtained in Ref. \cite{guslienko2008magnetic} are just semi-quantitative estimations.
	\begin{figure}
\includegraphics[width=86mm]{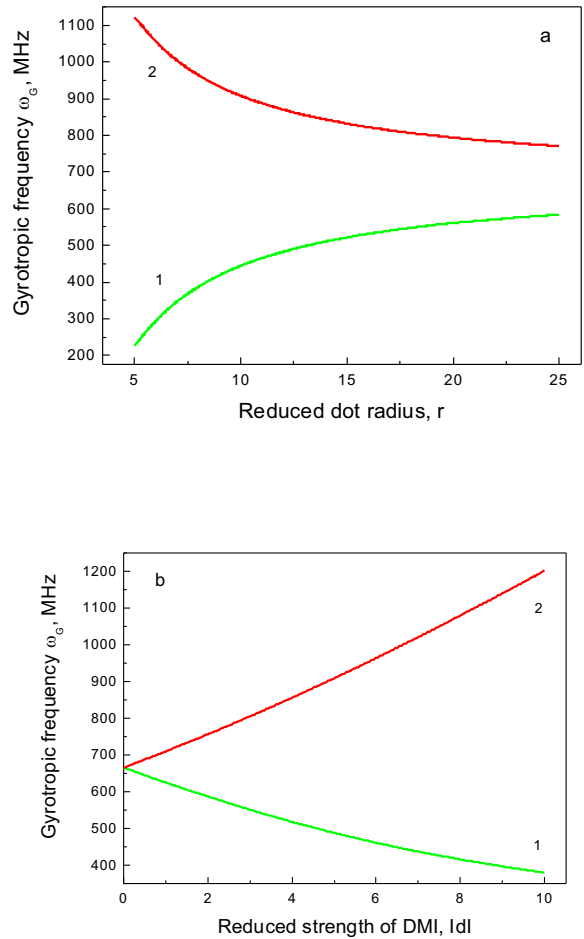}
\caption{\label{fig:Fig2} The gyrotropic frequencies of ultrathin skyrmion ground state circular dot calculated by Eq. (\ref{eq:A2}): (a) the frequencies vs. the reduced dot radius $r=R/L_e$,   $|d|=2.9$ (or $|D| =3 mJ/m^2$ using the magnetic parameters from Ref. \cite{buttner2015dynamics}); (b) the frequencies vs. the reduced DMI strength $|d|=|D|L_e/A$ at the reduced dot radius  $r=19$ (or  $R=275 nm$ as in Ref. \cite{buttner2015dynamics}). The solid green (1) and red (2) lines correspond to $d>0$  and $d<0$ ($d=-|d|<0$ for the skyrmion dot ground state). The parameters $M_s$ = 1190 $G$, $Q$=0.866 were taken from Ref. \cite{buttner2015dynamics}. The exchange length is  $L_e$=14.5 $nm$.}
\end{figure}

The gyrotropic frequency of the skyrmion state ultrathin dots given by Eq. (\ref{eq:A2}) is plotted in Figure 2. It increases/decreases with dot radius increasing for $d>0/d<0$  (Fig. 2a). However, the frequency at $d>0$ decreases at $R$ increasing if the side surface magnetostatic interaction for a moderate dot thickness is accounted. The eigenfrequency of the gyrotropic mode (\ref{eq:A2}) is splitted due to DMI in the chiral $\phi-$ and $\rho-$skyrmion state magnetic dots. There are two distinct gyrotropic frequencies for the skyrmions due to possible different signs of the skyrmion chirality $C$ of the dots in an array. The $m_{\rho}$ magnetization component plays a role of the chirality for $\rho-$skyrmions. The frequency splitting between the corresponding gyromodes is determined by the DMI strength $\Delta\omega_G(c)=\omega_M(L_e^2|d|/R)c^3/(1+c^2)^2$ and might be of several hundred $MHz$ (see Fig. 2b). However, the DMI energy for the skyrmion state dots is lower for such sign
 of chirality that the parameter $d=-|d|<0$ and corresponding gyrotropic frequency is higher (the curve 2 in Fig. 2a,b). The gyrofrequency (\ref{eq:A2}) is determined by the exchange stiffness $A$, intrinsic DMI parameter $D$, the perpendicular anisotropy constant $K$, as well as the dot saturation magnetization and radius $R$. The recent simulations by Zhang et al. \cite{zhang2015current} of the gyrotropic dynamics of the skyrmion in a free layer of circular nanopillar are in qualitative agreement with Eq. (\ref{eq:A2}). There is one gyrotropic frequency  $\omega_G(R)$, which rapidly decreases from 1.4 $GHz$ to zero with the dot radius $R$ increasing from 20 to 70 $nm$ for Co dot with the thickness 0.6 $nm$ (the magnetic parameters are the same as in Ref. \cite{sampaio2013nucleation}).\linebreak
\indent
Eq. (\ref{eq:A2}) for the gyrotropic frequency is applicable as well to 2D skyrmion triangular lattices in B20 compounds in the case of in-phase motions of the skyrmions in the lattice unit cells (the dot diameter $2R$ should be substituted to the skyrmion lattice period in this case). The eigenfrequency is about 1 $GHz$ for typical B20 skyrmion lattice parameters. 

We note that the skyrmion gyrotropic frequency in magnetic dots has not been observed yet. The reason for that is the stable skyrmions in ultrathin ferromagnetic dots were obtained very recently and there was no chance to detect the skyrmion excitations in such magnetic nanostructures. Another difficulty we foresee is large value of the magnetization damping (the resonance linewidth) in ultrathin Co films that does not allow applying a precise broadband ferromagnetic resonance technique for the gyrotropic frequencies detection in $\phi-$ and $\rho-$skyrmion state magnetic dots. The ultrathin Fe/Ni films \cite{chen2015room} with expected low  resonance linewidth are more promising to detect the skyrmion dynamics. This is a challenge for the future experiments.
\section{Conclusion}
We calculated low-frequency gyrotropic spin excitation modes of the skyrmion ground state cylindrical magnetic dots. The skyrmion was assumed to be stabilized at room temperature and zero external magnetic field due to interplay of the isotropic and Dzyaloshinskii-Moriya exchange interactions, perpendicular magnetic anisotropy and magnetostatic interaction. We considered the Bloch- and Neel-type magnetic skyrmions and assumed that the magnetic dot is thin enough (magnetization does not depend on the thickness coordinate). The skyrmion gyrotropic eigenfrequencies are calculated within the rigid skyrmion model as a function of the skyrmion equilibrium radius, dot radius and the dot magnetic parameters. Now the challenge is to detect experimentally the predicted gyrotropic eigenfrequncies.\\
\begin{acknowledgments}
K.G. acknowledges support by IKERBASQUE (the Basque Foundation for Science) and the Spanish MINECO grant MAT2013-47078-C2-1-P. Z.G. acknowledges support by the Russian Foundation for Basic Research (Grant No. 16-02-00336 A).
\end{acknowledgments}

\nocite{*}
\bibliography{references}
\bibliographystyle{apsrev4-1}

\end{document}